\newlength{\extralineskip}
\documentclass[12pt]{article}

\usepackage{epsfig}
\begin{document}
\begin{titlepage}
\begin{flushright}
          \begin{minipage}[t]{12em}
          \large UAB--FT--507\\
              January 2001                        
\end{minipage}
\end{flushright}
\vspace{\fill}

\vspace{\fill}

\begin{center}
\baselineskip=2.5em

{\large \bf 
 Pseudoscalar production in  
electromagnetic fields \\
by a Schwinger-like 
mechanism\footnote{Talk at QED2000, Trieste (October 2000)}}
\end{center}

\vspace{\fill}

\begin{center}
{\bf  Eduard Mass{\'o}}\\
\vspace{0.4cm}
     {\em Grup de F{\'\i}sica Te{\`o}rica and Institut de F{\'\i}sica
     d'Altes Energies\\
     Universitat Aut{\`o}noma de Barcelona\\
     08193 Bellaterra, Barcelona, Spain}
\end{center}
\vspace{\fill}

\begin{center}
\large Abstract
\end{center}
\begin{center}
\begin{minipage}[t]{36em}
In this talk I report on some recent calculations on
the production of pseudoscalars from intense electromagnetic fields.
\end{minipage}
\end{center}

\vspace{\fill}

\end{titlepage}

\clearpage

\addtolength{\baselineskip}{\extralineskip}

\section{Decay of classical background fields into pseudoscalars}

We would like to calculate the effective action for the 
background electromagnetic ${\bf E}$ and ${\bf B}$ fields,
\begin{equation}
e^{i S_{eff}[{\bf E}, {\bf B}]}
= \int {\cal D} \phi~
        e^{i S[\phi, {\bf E}, {\bf B}]}
\label{emasso:1}
\end{equation}
when integrating a pseudoscalar $\phi$ of mass $m$, that has an action
$S$ with a coupling
to the ${\bf E}$ and ${\bf B}$ fields of the form
\begin{equation}
S[ \phi, {\bf E}, {\bf B}]
= \int d^4 x~
        \frac{1}{2} \phi(x) 
        \left[ - \partial^2 - m^2 + f(x, {\bf E}, {\bf B})
        \right] \phi(x)
\label{emasso:2}
\end{equation}

Using the identities
\begin{eqnarray}
i \frac{\partial S_{eff}[{\bf E}, {\bf B}]}
        {\partial m^2}
&=& 
- \frac{ \int {\cal D} \phi~ \phi^2~
        e^{i S[\phi, {\bf E}, {\bf B}]} } 
        { \int {\cal D} \phi~ 
        e^{i S[\phi, {\bf E}, {\bf B}]} }
\nonumber \\
&=& - \frac{1}{2} \int d^4 x~
        G \left( x, x; {\bf E}, {\bf B} \right)
\nonumber \\
&=& - \frac{1}{2} \int d^4 x~
        \int \frac{d^4 p}{(2 \pi)^4}~
        G \left( p; {\bf E}, {\bf B} \right)
\label{emasso:3}
\end{eqnarray}
we can express 
the effective Lagrangian of the background fields 
in terms of the Green's function $G(p)$ of $\phi$ propagating 
in these fields.
\begin{equation}
{\cal L}_{eff} [{\bf E}, {\bf B}]
= \frac{i}{2}
        \int dm^2 \int \frac{d^4 p}{(2 \pi)^4}~
        G \left( p; {\bf E}, {\bf B} \right)
\label{emasso:4}
\end{equation}

Our objective now is to determine 1) $G(p)$ and 2) ${\cal L}_{eff}$.

{\bf 1)} The action (\ref{emasso:1}) contains the interaction Lagrangian 
\begin{equation}
{\cal L}_I (x) = {1\over 2} f(x) \phi^2(x)
\label{emasso:6}
\end{equation}
and leads to the following equation for the Green function
\begin{equation}
\left[
        \partial^2 + m^2 - f(x) \right]
G(x,0)
= \delta^4 (x)
\label{emasso:7}
\end{equation}

We will approximate  $f(x)$  by its
Taylor series near the reference point $x=0$ 
up to second order,
\begin{equation}
f(x)
= \alpha 
        + \beta_{\mu}  x^{\mu}
        + \gamma^2_{\mu \nu}  x ^{\mu} x^{\nu}
\label{emasso:8}
\end{equation}
The equation (\ref{emasso:7}) is then approximated by
\begin{equation}
\left[
        \partial^2 + m^2 - \alpha 
        - \beta_{\mu}  x^{\mu}
        - \gamma^2_{\mu \nu}  x ^{\mu} x^{\nu}
 \right]
G(x,0)
= \delta^4 (x)
\label{emasso:9}
\end{equation}
or, in momentum space,
\begin{equation}
\left[
- p^2 + m^2 - \alpha
        + i \beta_{\mu} \frac{\partial}{\partial p_{\mu}}
        + \gamma^2_{\mu \nu} 
                \frac{\partial}{\partial p_{\mu}}
                \frac{\partial}{\partial p_{\nu}}
\right]
G(p) = 1
\label{emasso:10}
\end{equation}

As is shown in detail in \cite{emasso:gmm},
the solution for $G(p)$ that satisfies the boundary conditions is  
\begin{equation}
G(p) 
= i \int_0^{\infty} ds~
        e^{- i s (m^2 - i \epsilon)}
        e^{i p_{\mu} A^{\mu \nu} p_{\nu} + B^{\mu} p_{\mu} + C}
\label{emasso:11}
\end{equation}
where
\begin{eqnarray}
\mbox{\boldmath A} &=&
\frac{1}{2}\, {\mbox{\boldmath $\gamma$}}^{-1} \cdot 
        \tan (2 {\mbox{\boldmath $\gamma$}} s)
\label{eqn15a}
\\
\mbox{\boldmath B} 
&=&
- \frac{i}{2}\, {\mbox{\boldmath $\gamma$}}^{-2} \cdot 
        \left[ 1 - \mbox{sec} (2 {\mbox{\boldmath $\gamma$}} s) \right] 
        \cdot \mbox{\boldmath $\beta$} 
\label{eqn15b}
\\
\mbox{C} 
&=&
i \alpha s - \frac{1}{2}\, \mbox{tr} 
        \left[ \ln \cos (2 {\mbox{\boldmath $\gamma$}} s) \right]
        + \frac{i}{8}\, \mbox{\boldmath $\beta$} \cdot 
                {\mbox{\boldmath $\gamma$}}^{-3} \cdot
                \left[ \tan (2 {\mbox{\boldmath $\gamma$}} s) 
                - 2 {\mbox{\boldmath $\gamma$}} s \right] 
                \cdot \mbox{\boldmath $\beta$}
\label{emasso:12}
\end{eqnarray}

{\bf 2)} The effective Lagrangian is obtained by substituting 
$G(p)$ in (\ref{emasso:4}) and carrying out the integration over $m^2$,
\begin{equation}
{\cal L}_{eff}
= - \frac{i}{2}
        \int_0^{\infty} \frac{ds}{s}
        \int \frac{d^4 p}{(2 \pi)^4}~
        \exp \left\{- i s m^2 + i \mbox{\boldmath p} \cdot
                        \mbox{\boldmath A} \cdot
                        \mbox{\boldmath p}
                        + \mbox{\boldmath B} \cdot
                        \mbox{\boldmath p} 
                        + \mbox{C} \right\}
\end{equation}
After evaluation of the Gaussian integral  we finally get
\begin{equation}
{\cal L}_{eff}
= - \frac{1}{32 \pi^2} 
        \int_0^{\infty} \frac{ds}{s^3}~
        e^{- i s (m^2 - \alpha)}
        \left[
           \mbox{det} \left( 
                \frac{2 {\mbox{\boldmath $\gamma$}} s}
                {\sin 2 {\mbox{\boldmath $\gamma$}} s} 
                \right)
        \right]^{\frac{1}{2}}
        e^{i l(s)}
\label{emasso:13}
\end{equation}
where
\begin{equation}
l(s)
= \frac{1}{4}\, \mbox{\boldmath $\beta$} \cdot 
        {\mbox{\boldmath $\gamma$}}^{-3} \cdot
        \left[ \tan ({\mbox{\boldmath $\gamma$}} s) 
        - {\mbox{\boldmath $\gamma$}} s \right] \cdot
        \mbox{\boldmath $\beta$}
 \label{emasso:14}
\end{equation}
When the effective Lagrangian has an imaginary part, 
there is particle production with a probability density given by 
\begin{equation}
w=  2~\mbox{Im}  {\cal L}_{eff}[{\bf E}, {\bf B}]
\label{emasso:5}
\end{equation}
In turn, a
non-zero value for $\mbox{Im}  {\cal L}_{eff}$ may arise depending on the
sign of the $\mbox{\boldmath $\gamma$}$-matrix eigenvalues.

\section{Effective $F^2\phi^2$ interactions}

We assume the standard pseudoscalar-two photon coupling
\begin{equation}
{\cal L}_{\phi\gamma\gamma} = \frac{1}{8} g \phi
        \epsilon^{\mu \nu \rho \sigma} {F}_{\mu \nu}
       {F}_{\rho \sigma}
\label{emasso:last}
\end{equation}

We shall now show that, for the purposes of calculating ${\cal L}_{eff}$
of  external ${\bf E}$ and ${\bf B}$, we can use an interaction Lagrangian 
of the type displayed in (\ref{emasso:6}).

We first calculate the two-photon two-pseudoscalar amplitude
in momentum space,
\begin{equation}
 4 \left( \frac{1}{4} g \widetilde{\phi} \right)^2
\epsilon^{\mu \nu \rho \sigma} k_{\mu} \widetilde{F}_{\rho \sigma}
\frac{- i g_{\nu \nu^{\prime}}}{k^2}
\epsilon^{\mu^{\prime} \nu^{\prime} \rho^{\prime} \sigma^{\prime}} 
(-k_{\mu^{\prime}})
\widetilde{F}_{\rho^{\prime} \sigma^{\prime}}
\label{emasso:15}
\end{equation}
Due to the presence of the $k^2$ term in the denominator, the effective
coupling (\ref{emasso:15}) is non-local. However, when we calculate the 
effective action for the external 
electromagnetic field the momentum $k$ is integrated over. 
One can therefore make use of the identity 
\begin{equation}
\int d^4 k~ k_{\mu} k_{\mu^{\prime}}\, g( k^2 )
= \int d^4 k~
         \frac{ g_{\mu \mu^{\prime}}k^2}{4}\,  g(k^2) 
\label{emasso:16}
\end{equation}
to simplify (\ref{emasso:15}). Thus, we can reduce the effective 
two-photon two-pseudoscalar to a local interaction vertex.  
Back in configuration space, it is given by
 \begin{equation}
{\cal L}_{I}
= - \frac{1}{4} g^2 \phi^2
        {F}_{\mu \nu}
        {F}^{\mu \nu} = \frac{1}{2} g^2 \phi^2
        ( {\bf{E}}^2 - {\bf{B}}^2 )
\label{emasso:17}
\end{equation}
so that we can identify $f(x)$ in (\ref{emasso:6}) with
\begin{equation}
f(x) =  g^2   ( {\bf{E}}^2 - {\bf{B}}^2 )
\label{emasso:18}
\end{equation}

%-----------------------------------------------------------------
\section{Pseudoscalar production in a cylindrical capacitor}

In order to have a non trivial ${\cal L}_{eff}$, one needs non-zero
second derivatives of the electromagnetic fields as they appear in
expression  (\ref{emasso:18}), which imply a non-zero 
$\mbox{\boldmath $\gamma$}$-matrix. 
We illustrate it in the simple situation of the electric field
inside a cylindrical capacitor.

The modulus of the electric field inside a cylindrical capacitor whose
axis lies along 
the $z$-axis depends only on $\rho=(x^2+y^2)^\frac{1}{2}$,
\begin{equation}
E(\rho)= \frac{\lambda}{2 \pi} \frac{1}{\rho}
\label{emasso:19}
\end{equation} 
with $\lambda$ the linear electric charge density. It follows that
\begin{equation}
f(\rho)= g_c^2 \left( \frac{1}{\rho^2} \right)
\label{emasso:20}
\end{equation}
where $g_c \equiv \lambda g/2 \pi$

Expanding the fields near some reference point $(x_0,y_0,z_0)$ 
with $\rho_0=(x_0^2+y_0^2)^\frac{1}{2}$, we identify
\begin{equation}
\alpha =  
 \frac{g_c^2}{\rho_0^2} 
\label{emasso:21}
\end{equation}
\begin{equation}
(\mbox{\boldmath $\beta$})_i
= - \frac{2 g_c^2}{\rho_0^4}
        (x_0, y_0)
\label{emasso:22}
\end{equation}
and
\begin{equation}
(\mbox{\boldmath $\gamma^2$})_{ij}
= \frac{g_c^2}{\rho_0^6}
\left(
\begin{array}{cc}
-y_0^2 + 3 x_0^2 & 4 x_0 y_0  \\
4 x_0 y_0 & -x_0^2  + 3 y_0^2
\end{array}
\right)
\label{emasso:23}
\end{equation}
(We only include the $x-y$ entries). 
In diagonal form $\mbox{\boldmath $\gamma^2$}$ reads
\begin{equation}
({\mbox{\boldmath $\gamma$}}^2_D)_{i j} 
= \frac{g_c^2}{\rho_0^4}
\left(  
\begin{array}{cc}
3 & 0  \\
0 & -1
\end{array}
\right)
\equiv
\left(
\begin{array}{cc}
  a^2  & 0 \\
0 & - b^2 
\end{array}
\right)
\label{emasso:24}
\end{equation}

Introducing $\alpha$, $\mbox{\boldmath $\beta$}$ and
 $\mbox{\boldmath $\gamma$}$ in ${\cal L}_{eff}$ as given 
in (\ref{emasso:13}), we get the following expression 
\begin{equation}
{\cal L}_{eff}
= - \frac{1}{32 \pi^2} 
        \int_0^{\infty} \frac{ds}{s^3}~
        e^{- i s (m^2 - \alpha)}
        \sqrt{
            \frac{2 a s}{\sinh 2 a s}
        }~
\sqrt{
        \frac{2 b s}{\sin 2 b s}    
        }~ 
       e^{i l(s)}
\label{emasso:26}
\end{equation}
where
\begin{equation}
l(s)
=
    \lambda   \left( a s - \tanh a s \right) \\
\lambda = g_c^4 \rho_0^{-6} a^{-3} = 
\frac{g_c}{3 \, \sqrt{3}}
\label{emasso:2728}
\end{equation}

One can perform the integration in (\ref{emasso:26})
by extending $s$ to the complex plane. The details of the integration 
can be found in \cite{emasso:gmm}, where it is found that
\begin{equation}
\mbox{Im}~ {\cal L}_{eff}
=
\frac{a^\frac{3}{2} b^\frac{1}{2}}{8 \pi^2}
        \sum_{n=0}^{\infty}
        (-1)^n \, C_n \,
        e^{- \chi (2n + 1) \pi}
\label{emasso:29}
\end{equation}
\begin{equation}
C_n
=
\int_0^{\pi} du~
              \frac{
                e^{- \chi u}
                e^{- \lambda \cot (u/2)}
              }{  
                \left[ u + (2n + 1) \pi \right]^{2}
                \left[ \sin u \right]^{\frac{1}{2}}
                \left[\sinh \left( 
                            \frac{b}{a}
                            \left[ u + (2n + 1) \pi \right]
                         \right)\right]^{\frac{1}{2}}
               } 
\label{emasso:30} 
\end{equation} 
(we can put $b/a=1/ \sqrt{3}$). We have defined
\begin{equation}
\chi =  \frac{m^2 - \alpha}{2 a} +  \frac{\lambda}{2}
=  \frac{1}{\sqrt{3}} \left(
\frac{1}{2}  \frac{m^2 \rho_0^2}{ g_c}
-  \frac{1}{ 3}\, g_c \right)
\label{emasso:31}     
\end{equation}

Finally, the expression for the probability per unit volume and per
unit time for pseudoscalar production inside a cylindrical capacitor
is
\begin{equation}
w =
\frac{3^\frac{3}{4}}{4 \pi^2}
       \frac{g_c^2}{\rho_0^4} 
         C_0 \,
        e^{- \chi  \pi}
\end{equation}
were we kept only the leading $n = 0$ term in (\ref{emasso:29}).

\vspace{1cm}

Work partially supported by the CICYT Research Project AEN99-0766.

\end{document}